\documentclass[runningheads]{llncs}
\usepackage[T1]{fontenc}
\usepackage{graphicx}
\usepackage{amsmath}
\usepackage{amssymb}
\usepackage{booktabs}
\usepackage{multirow}
\usepackage[pagebackref,breaklinks,colorlinks]{hyperref}
\usepackage{pifont}

\usepackage{color}

\newcommand{\figref}[1]{Fig.~\ref{fig:#1}}
\newcommand{\tabref}[1]{Table~\ref{tab:#1}}

\newcommand{\ra}[1]{\renewcommand{\arraystretch}{#1}} 
\newcommand{\norm}[1]{\left\lVert#1\right\rVert}

\newcommand{\demo}{
\begin{figure}[t!]
    \centering
    \includegraphics[width=\linewidth]{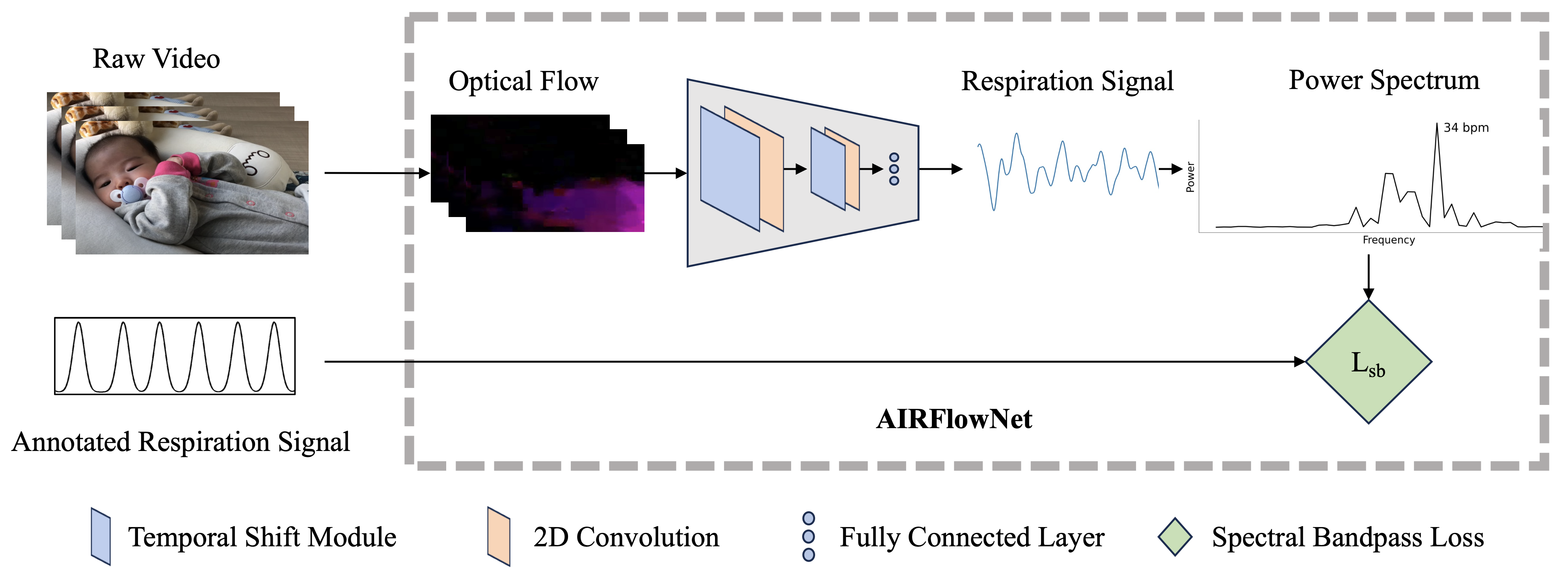}
    \caption{Video-based respiration signal and rate estimation from our AIRFlowNet model trained with spectral bandpass loss and tested on our annotated infant respiration (AIR-125) dataset.}
    \label{fig:dmem}
\end{figure}
}

\newcommand{\datasample}{
\begin{figure*}[t!]
    \centering
    \includegraphics[width=\linewidth]{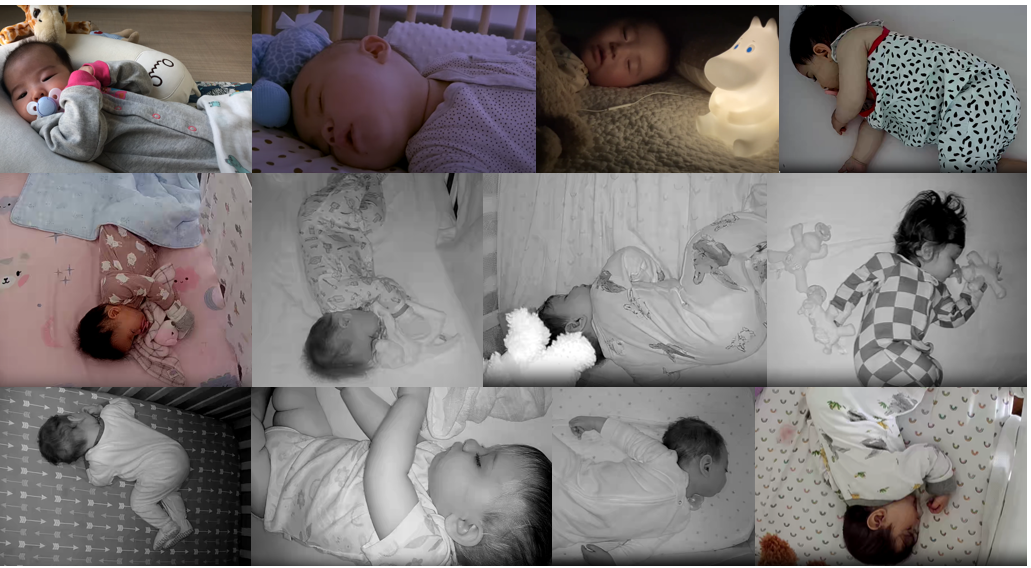}
    \caption{Sample frames in our annotated infant respiration (AIR-125) dataset showing the diversity in pose, illumination, background, and camera type.}
    \label{fig:dataset-sample}
\end{figure*}
}

\newcommand{\datadist}{
\begin{figure}[t!]
    \centering
    \includegraphics[width=\linewidth]{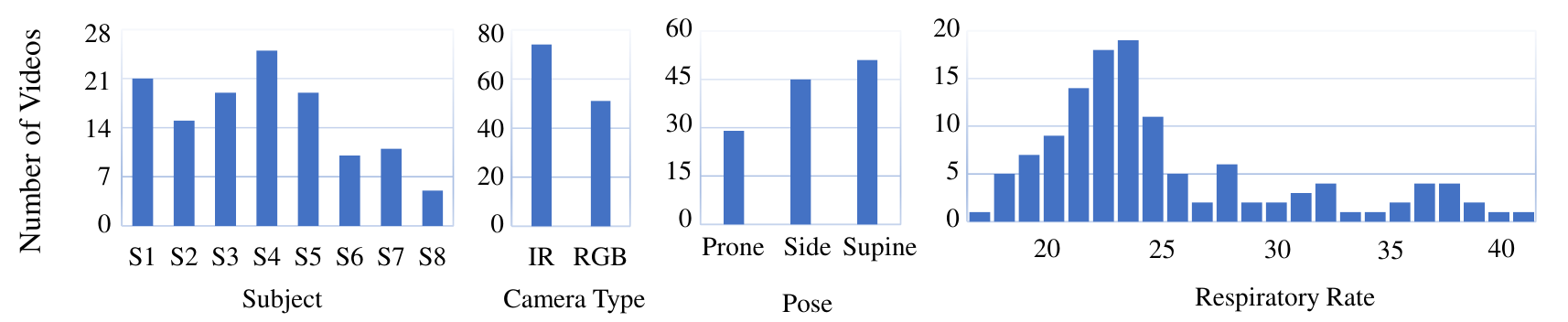}
    \caption{Distribution of pose, camera type per subject, and respiratory rates (breaths per minute) in our annotated infant respiration (AIR-125) dataset.} 
    \label{fig:datadist}
\end{figure}
}

\newcommand{\publicdata}{
\begin{table*}[t!]
    \centering
        \caption{Datasets and methods in the literature for respiration estimation. PPG: Photoplethysmogram, PR: Pulse Rate, RR: Respiratory Rate, Resp: Respiration waveform, AU: Action Unit, BVP: Blood Volume Pulse, EEG: Electroencephalogram, ECG: Electrocardiogram, SpO2: Blood oxygenation, DL: Deep Learning, SP: Signal Processing.}
    \resizebox{\columnwidth}{!}{
    \begin{tabular}{lllrrcl}
        \toprule
         Dataset & Ground Truth & Domain & Videos & Subjects &  Public & Method \\
         \midrule
         SCAMPS \cite{mcduff2022scamps} & PPG, PR, RR, Resp, AU & Adult & 2800 & 2800 & \ding{51} & DL \\ 
         COHFACE \cite{heusch2017reproducible} & Resp, BVP & Adult & 160 & 40 & \ding{51} & SP \\
         MAHNOB \cite{soleymani2011multimodal} & ECG, EEG, Resp & Adult & 527 & 27 & \ding{51} & None \\
         AFRL \cite{estepp2014recovering} & ECG, EEG, PPG, PR, RR & Adult & 300 & 25 & \ding{55} & SP \\
         OBF \cite{li2018obf} & RR, PPG, ECG & Adult & 212 & 106 & \ding{55} & SP \\
         \midrule
         Villarroel \textit{et al.} \cite{villarroel_non-contact_2019} & Resp, PPG, SpO2 & Infant & 384 & 30 & \ding{55} & DL \\
         Földesy \textit{et al.} \cite{foldesy2020reference} & Resp & Infant & 1440 & 7 & \ding{55} & DL \\
         Kyrollos \textit{et al.} \cite{kyrollos2021noncontact} & Resp & Infant & 20 & 1 & \ding{55} & SP \\
         Lorato \textit{et al.} \cite{lorato2021towards} & Resp & Infant & 90 & 2 & \ding{55} & SP \\
         Tveit \textit{et al.} \cite{tveit2016motion} & RR & Infant & 6 & 2 & \ding{55} & SP \\
         \midrule
         \textbf{AIR-125 (ours)} & Resp, RR, Pose & Infant & 125 & 8 & \ding{51} & DL \\
         \bottomrule
    \end{tabular}}

    \label{tab:dataset-table}
\end{table*}
}

\newcommand{\tabcohface}{
\begin{table*}[t!]
    \centering
    \caption{Comparison of different motion-based ($\dagger$) and color-based ($\star$) methods under various train $\to$ test data configurations (COHFACE for adult, AIR-125 for infant). Note that \cite{guo2021remote} does not learn from respiration data.}
    \resizebox{\columnwidth}{!}{
    \ra{0.8} 
    \begin{tabular}{lrrrrrrrrrrr}
        \toprule
         & \multicolumn{3}{c}{Adult $\to$ Adult} & \phantom{a} & \multicolumn{3}{c}{Adult $\to$ Infant} & \phantom{a} & \multicolumn{3}{c}{Infant $\to$ Infant}  \\
         \cmidrule[0.5pt]{2-4} \cmidrule[0.5pt]{6-8} \cmidrule[0.5pt]{10-12}
        & MAE  $\downarrow$ & RMSE $\downarrow$ & $\rho \uparrow$ & & MAE  $\downarrow$ & RMSE $\downarrow$ &  $\rho \uparrow$ & & MAE  $\downarrow$ & RMSE $\downarrow$ & $\rho \uparrow$ \\
        Method   & (bpm) & (bpm) &  & & (bpm) & (bpm) & &  & (bpm) & (bpm) &  \\
         \midrule
         Eff-Phys$^\star$ \cite{liu2023efficientphys} & 4.07 & 5.46 &  0.27 & & 7.21 & 9.08 & 0.40 &  & 6.22 & 7.83 & 0.44 \\
         DeepPhys$^\star$ \cite{chen2018deepphys} & 2.68 & 4.64 & 0.36 & & 6.76 & 9.29 & 0.38 &  & 6.06 & 8.79 & 0.38 \\
         TS-CAN$^\star$ \cite{liu2020multi} & 2.25 & 3.96 & 0.53 & & 8.84 & 11.6 & 0.20 & & 6.35 & 7.54 & 0.50\\
         Guo \textit{et al.}$^\dagger$ \cite{guo2021remote}  & 1.04 & 2.45 & \textbf{0.82} & & 4.68 & 6.74 & 0.32 &  & 4.68 & 6.74 & 0.32 \\
         \textbf{AIRFlowNet}$^\dagger$ & \textbf{1.01} & \textbf{2.20} & 0.76 & & \textbf{4.16} & \textbf{5.98} & \textbf{0.41} &  & \textbf{2.91} & \textbf{5.40} & \textbf{0.72} \\
         \bottomrule
    \end{tabular}}
    \label{tab:cohface}
\end{table*}
}

\newcommand{\tabablation}{
\begin{table*}[t!]
    \centering
    \caption{AIRFlowNet performance when trained and tested on AIR-125 data using common loss functions and our novel spectral bandpass loss $L_\text{sb}$.}
    \ra{0.9} 
    \begin{tabular}{lcrrrrrrr}
        \toprule
        Metric & \phantom{aaaaa} & $L^1$ & \phantom{a} & $L^2$ & \phantom{a} & $-\rho$ & \phantom{a} & $L_\text{sb}$ (ours) \\
        \midrule
        MAE & $\downarrow$ & 3.49 & & 3.46 &  & 2.96 &  & \textbf{2.91} \\
        RMSE & $\downarrow$ & 6.26 &  & 5.36 &  & \textbf{5.34} &  & 5.40 \\
        $\rho$ & $\uparrow$ & 0.64 &  & 0.70 &  & 0.71 &  & \textbf{0.72} \\
        \bottomrule
    \end{tabular}
    \label{tab:ablation}
\end{table*}
}

\begin{document}

\title{Automatic Infant Respiration Estimation from Video: A Deep Flow-based Algorithm and a Novel Public Benchmark}
\titlerunning{Automatic Infant Respiration Estimation}
%
%
%
%

\author{Sai Kumar Reddy Manne \inst{1,2}, Shaotong Zhu\inst{2},\\ Sarah Ostadabbas\inst{2}, Michael Wan\inst{1,2}$^*$}

\authorrunning{Sai Kumar Reddy Manne et al.}
\institute{Roux Institute, Northeastern University, Portland ME, USA\\
\and
Augmented Cognition Lab, Department of Electrical \& Computer Engineering\\Northeastern University, Boston MA, USA\\
$^*$Corresponding author: \email{mi.wan@northeastern.edu}
}
\maketitle              
%

\begin{abstract}
Respiration is a critical vital sign for infants, and continuous respiratory monitoring is particularly important for newborns. However, neonates are sensitive and contact-based sensors present challenges in comfort, hygiene, and skin health, especially for preterm babies. As a step toward fully automatic, continuous, and contactless respiratory monitoring, we develop a deep-learning method for estimating respiratory rate and waveform from plain video footage in natural settings. Our automated infant respiration flow-based network (AIRFlowNet) combines video-extracted optical flow input and spatiotemporal convolutional processing tuned to the infant domain. We support our model with the first public annotated infant respiration dataset with 125 videos (AIR-125), drawn from eight infant subjects, set varied pose, lighting, and camera conditions. We include manual respiration annotations and optimize AIRFlowNet training on them using a novel spectral bandpass loss function. When trained and tested on the AIR-125 infant data, our method significantly outperforms other state-of-the-art methods in respiratory rate estimation, achieving a mean absolute error of $\sim$2.9 breaths per minute, compared to $\sim$4.7--6.2 for other public models designed for adult subjects and more uniform environments\footnote{Our code and the manually annotated NNS in-the-wild dataset can be found at \url{https://github.com/ostadabbas/Infant-Respiration-Estimation}. Supported by MathWorks and NSF-CAREER Grant \#2143882.}.

\end{abstract}

\section{Introduction}
\label{sec:intro}
\demo
From an infant's first breath in the seconds after birth, respiration becomes a critical vital sign in early life, with irregularities revealing complications ranging from apnea and respiratory distress in neonates \cite{reuter2014respiratory}, to  respiratory syncytial virus (RSV) infection in months-old infants leading in severe cases to bronchiolitis or pneumonia \cite{hall2009rsv}, or even potentially to sudden infant death syndrome (SIDS) \cite{THACH2005343}. Continuous respiratory monitoring is particularly important for preterm infants during \textit{ex utero} development in neonatal intensive care units (NICUs), where contactless sensors are desirable for comfort and hygiene, and to prevent skin damage, during this sensitive period \cite{villarroel_non-contact_2019}. We present a novel vision-based deep learning method for detecting an infant's respiration waveform and respiratory rate (see \figref{dmem}), as a step toward automated monitoring of infant breathing for both everyday and clinical diagnostic use. Only a few papers have explored deep learning-based infant respiration estimation (see \tabref{dataset-table}), and due to logistical and privacy constraints on infant data collection, none of them publish their data or models and many draw data from just one or two infant subjects \cite{tveit2016motion,kyrollos2021noncontact,lorato2021towards,foldesy2020reference}. As part of this work, we publish the \textit{\textbf{a}nnotated \textbf{i}nfant \textbf{r}espiration dataset of \textbf{125} videos}, \textbf{AIR-125}, with ground truth respiratory rates and waveforms from eight infant subjects, to support public and reproducible research. AIR-125 features minute-long videos sourced from baby monitors and smartphone cameras in natural infant settings, with varying illumination, infant poses, age groups, and respiratory rates. We use manual respiration annotations rather than sensor-captured ground truth to enable data collection from various sources, but also include synthetically-generated respiration waveforms to maintain compatibility with existing models and benchmarks.

\publicdata

Existing approaches for respiration measurement \cite{guo2021remote,liu2023efficientphys} track motion using optical flow or track subtle color changes in skin pixels. The flow-based methods are prone to errors from noise or subject motion and the color-based methods rely on visible skin pixels in the video, which may be scarce for infants who are heavily covered or sleeping in an awkward pose like those in AIR-125. Hence, we also propose a new model, the \textit{\textbf{a}utomated \textbf{i}nfant \textbf{r}espiration \textbf{flow}-based network}, \textbf{AIRFlowNet}, which learns to isolate the periodic respiratory motion in a noisy environment without the need for visible skin in the video. Current respiration models are trained with ground truth obtained from contact sensors perfectly synchronized with videos \cite{chen2018deepphys,liu2020multi,liu2023efficientphys}. We introduce a novel \textit{spectral bandpass loss function} which encourages alignment in the frequency domain while forgiving absolute temporal shifts, enabling more effective training with our manual annotations. When trained and tested on AIR-125 infant data, AIRFlowNet significantly outperforms other state-of-the-art respiration models. 


In sum, our key contributions include (1) the first public annotated infant respiration dataset (AIR-125), (2) a motion-based infant respiration estimation model (AIRFlowNet) with a novel spectral bandpass loss achieving best-in-class performance, and (3) performance comparison of public color- and motion-based respiration models on infant and adult datasets.

\section{Related Work}
\label{sec:related_work}


Respiration induces cyclical expansion and contraction in the chest and abdomen regions. 
Motion-based methods track this subtle motion in videos to estimate the respiration signal. Tveit \textit{et al.} \cite{tveit2016motion} use Rietz transform in a phase-based algorithm to track respiratory patterns and test their model on infant and adult subjects.
Shao \textit{et al.} \cite{shao2014noncontact} estimate both heart rate and breathing rate simultaneously by tracking shoulder motion and color changes in a subject's face. Guo \textit{et al.} \cite{guo2021remote} improve the motion tracking using optical flow and human segmentation from pretrained deep learning models. Lorato \textit{et al.} \cite{lorato2021towards} use a two-stage approach to detect and reject video clips with severe motion, followed by a handcrafted feature-based rate estimation. Kyrollos \textit{et al.} \cite{kyrollos2021noncontact} use depth information along with RGB videos to improve the accuracy. Földesy \textit{et al.} \cite{foldesy2020reference} propose an incremental learning model to extract accurate frequency from a noisy estimate.

Another common approach for respiration estimation is based on the complex photoplethysmography (PPG) signal, a superposition of the slowly changing DC respiration component and the rapidly changing AC pulse component. Respiration waveform estimation based on color tracking was first introduced in DeepPhys \cite{chen2018deepphys}. 
Temporal shift modules were introduced in \cite{liu2020multi}, in place of computationally expensive 3D convolutions, to improve efficiency of the model. 
Villarroel \textit{et al.} \cite{villarroel_non-contact_2019} present a PPG signal extraction method for continuous infant monitoring in NICU setting. They use a multi-task CNN for segmenting skin pixels and detecting the presence of a subject in the camera, followed by simple pulse and breathing estimation.
In \cite{liu2021metaphys}, a few-shot adaptation of the base temporal shift deep learning model is used to improve results for individual subjects. In \cite{yu2019remote}, a novel loss based on Pearson correlation is used to train the model, improving the estimation accuracy compared to a model trained with $L^1$ or $L^2$ losses. To the best of our knowledge, our work provides the first comparison study between color- and motion-based approaches for infant subjects, complementing the one existing study for adult subjects \cite{wang2022camera}.

Finally, thermal imaging can be used to track the alternating cold and warm air flowing from the nose during inhalation and exhalation. Such methods \cite{hochhausen2018estimating,jakkaew2020non,lorato2021towards} usually track a region of interest (ROI) in the nasal area across the video. 
Thermal cameras can be used in complete darkness, making them a good alternative to RGB cameras, but their setup cost prevents ubiquitous deployment. 

\datasample

\section{AIR-125: An Annotated Infant Respiration Dataset}
\label{sec:infant-data}




Available physiological measurement datasets are created synthetically \cite{mcduff2022scamps} or extract reference physiological signals from contact-based systems and require the subjects situated in a controlled environment \cite{heusch2017reproducible,villarroel_non-contact_2019}. Unlike the existing datasets, AIR-125 features infant videos collected from a range of sources and settings to enable training and testing of flexible models useful for monitoring in everyday settings outside of the lab; our manual annotation process makes broad collection easier by eliminating equipment and recruitment constraints.
Our primary source is baby monitor footage from five infants, captured during daylight and nighttime sleep sessions in-crib by our clinical team under Institutional Review Board (IRB \#22-11-32) approval, with no other constraints on pose, lighting, clothing, and face visibility. The monitor switches between RGB and infrared (IR) modes depending on the light. For further diversity, we also source clips from three infant subjects on YouTube. From both sources, we extract a combined dataset of 125 videos, each approximately 60 seconds long. For respiration annotations, we parse through video frames, focusing on thoracic or abdominal motion to determine start times of exhalation cycles aided by the VGG Image Annotator \cite{dutta2019vgg}. 
Annotated respiratory rates range from 18--42 breaths per minute; see \figref{datadist} for distributions by subject, pose, camera type, and respiratory rate.


\datadist

The annotations from each video clip are converted to an impulse sequence, with one pulse per exhalation start time label. To create smooth waveforms that are analogous to  signals from a contact-based respiration system, the impulse sequence is Gaussian filtered with an empirically determined radius of $4$ frames. The smoothed waveform is used as ground truth signal for our respiratory rate estimation methods. The video resolutions range from $854\times480$ to $1920\times1080$ and frame rates from 10--30 Hz. 

\section{Methodology}
\label{sec:motion_dl}

\subsection{AirFlowNet Architecture}
Color-based approaches track imperceptible color changes to estimate the remote photoplethysmography (rPPG) signal and isolate the breathing signal from a complex superposition of other vital signals \cite{chen2018deepphys}. Hence, these methods are prone to errors unless severe restrictions are imposed on the environment such as constant illumination, still subjects, and no camera motion. On the other hand, existing motion-based approaches use hand-crafted features, classical computer vision techniques, or pretrained deep learning models to track specific regions of interest to estimate breathing signals \cite{koolen2015automated,tveit2016motion,shao2014noncontact}. To alleviate the shortcomings of these two approaches, we propose our annotated infant respiration flow-based network (AIRFlowNet), depicted in \figref{dmem}, which processes optical flow input with a spatio-temporal convolutional network and isolates a clean respiration signal from a noisy video with possible subject or camera motion. Using optical flow input also eliminates the need to retrain a model when testing on videos from different camera types such as RGB and IR cameras. 



We use a simple yet accurate implementation of coarse-to-fine optical flow \cite{liu2009beyond} for our experiments. The optical flow is generated at $96\times 96$ resolution to preserve the subtle motion induced by respiration and reduce the effects of spatial noise in the flow calculation. The calculated flow vectors are stored in HSV color space at a frame rate of $5$ Hz. 


We base our convolutional network on EfficientPhys \cite{liu2023efficientphys}, adapting it to optical flow inputs. We replace the difference layer in EfficientPhys with a convolution layer followed by a batch-normalization layer as our inputs are Z-score normalized in the preprocessing stage. The first convolution layer follows a series of temporal shift modules \cite{liu2020multi} and convolution layers that efficiently compute temporal features by shifting the channels across time dimension. Self-attention blocks following the temporal shift modules refine the features to appropriately weigh different spatial locations that correspond to respiration motion. A dense layer is used at the end of the network to estimate a 1D respiration signal. Unlike EfficientPhys, which estimates the first order derivative of the signal, our model estimates the respiration signal directly.

    
\subsection{Spectral Bandpass Loss}
\label{sec:psd_mse}
Current respiration estimation models train the networks using the $L^2$ loss between the ground truth signal and the predicted respiration waveform. While $L^2$ loss is useful for training with a ground truth signal that is precisely synchronized with the video, such as that obtained from electronic sensors, any temporal misalignment can lead to erroneous results. Since our manual annotations do not enjoy near-perfect alignment, we employ a new loss function that imposes a penalty entirely in the frequency domain, to prevent slight temporal misalignments from impeding effective learning of respiratory rate.

For any waveform $x=x(t)$, we use the fast Fourier transform $\mathcal{F}$ to define its corresponding power spectral density $X_\text{PSD} := |\mathcal{F}(x-x_0)|^2$, where $x_0$ is the temporal mean of $x$. After computing power spectral densities $Y_\text{PSD}$ and $\hat{Y}_\text{PSD}$ for the predicted ($y$) and the reference waveform ($\hat{y}$) respectively, we filter out the power from frequencies outside the normal infant breathing range of 0.3--1.0 Hz using a bandpass filter, $B(\cdot)$. The filtered power spectral densities are normalized to have a unit cumulative power. 
We define the \textbf{spectral bandpass loss $L_\text{sb}$} between $y$ and $\hat{y}$ by 
\begin{equation}
    L_\text{sb}(y,\hat{y}) = \norm{\frac{B(Y_\text{PSD})}{\sum_{\xi \in \Xi}B(Y_\text{PSD})} - \frac{B(\hat{Y}_\text{PSD})}{\sum_{\xi \in \Xi}B(\hat{Y}_\text{PSD})}}_2,
\end{equation}
with the outer norm being the $L^2$ norm, and $\Xi$ constituting the set of frequencies in the power spectrum.


\section{Evaluation and results}
\label{sec:exp-res}

\subsection{Experimental Setup}
\subsubsection{Datasets}
We evaluate our model on a public adult dataset, COHFACE \cite{heusch2017reproducible}, along with our infant dataset. COHFACE contains 160 webcam clips from 40 subjects, each approximately 60 seconds long. The videos are recorded at $640\times480$ resolution and 20 Hz, under both ambient and normal lighting conditions. Reference respiration signals come from a respiration belt readout at 32 Hz. 

\subsubsection{Training}
To train and evaluate our models on COHFACE, we use the rPPG-toolbox \cite{liu2022deep}, which provides a training framework for several physiological signal estimation models designed for adult subjects. We use the toolbox to train current state-of-the-art physiological measurement models: DeepPhys \cite{chen2018deepphys}, TS-CAN \cite{liu2020multi}, and EfficientPhys \cite{liu2023efficientphys}. For the color-based model training, the dataset is preprocessed to detect a face in each frame using Haar cascade classifier. The frames are then cropped around the face bounding box, and resized to a lower resolution of $96\times96$. All the models are trained with $L^2$ loss to generate a continuous signal for each clip. To train AIRFlowNet, we estimate the optical flow for each video and do not perform any face-based preprocessing. The rest of the training methodology is identical between all the trained models.

\subsubsection{Post-processing}
The estimated respiration signals are first filtered using a bandpass filter to remove noise from external sources. The lower and upper cut-off frequencies for the bandpass filter are $0.3$ Hz and $1.0$ Hz, covering the normal infant respiratory rates of 18--60 breaths per minute. The filtered signal is then transformed to frequency domain through a fast Fourier transform. We perform power spectral density analysis to determine the frequency with the maximal power as the predicted respiratory rate. We calculate three metrics that are commonly used in the literature to compare the different approaches: mean absolute error (MAE), root mean squared error (RMSE), and Pearson's correlation coefficient ($\rho$).



\subsection{Results and Analysis}
\label{sec:results}

\tabcohface
\tabablation

We tabulate results from the following three experimental configurations (training dataset $\to$ testing dataset) in \tabref{cohface}.

\subsubsection{Adult $\to$ Adult}
We compare our model with color-based methods \cite{chen2018deepphys,liu2020multi,liu2023efficientphys} trained on COHFACE and the motion-based method from \cite{guo2021remote}. Since COHFACE has very still subjects with no external motion, motion-based approaches perform better than color-based models. 


\subsubsection{Adult $\to$ Infant}
We quantify the domain generalizability of all approaches by training on COHFACE and testing on AIR-125. The AIR-125 dataset is divided into train and test splits, with 50 clips from 3 subjects in the training split, and 75 videos from the remaining 5 subjects in the test split. 
Our model demonstrates better generalizability as it is agnostic to camera type and brightness changes owing to the optical flow input.

\subsubsection{Infant $\to$ Infant}
For a fair comparison, we train and test both AIRFlowNet and the other models designed for adult subjects purely on AIR-125 data. Even when trained on infant data, the other models struggle to attain acceptable performance, exhibiting high mean absolute error and low Pearson's correlation. Our model achieves the best infant-domain performance, and the quantitative results even rival the performances of adult models tested on adult data.


\subsubsection{Ablation Study}
To demonstrate the effectiveness of our spectral bandpass loss $L_\text{sb}$, we compare results of AIRFlowNet trained and tested on AIR-125 data under $L_\text{sb}$ and three other common loss functions---$L^1$, $L^2$, and negative Pearson loss ($-\rho$) \cite{yu2019remote}---in \tabref{ablation}. The $L_\text{sb}$ loss performs best, but $-\rho$ is also effective, likely because it also relaxes constraints on strictly matching the ground-truth signal, compared to the $L^1$ and $L^2$ losses. Note, however, that training with $-\rho$ requires ground truth waveforms (synthetically generated in AIR-125), whereas $L_\text{sb}$ can be trained with respiration exhalation timestamps alone.


\section{Conclusion}
\label{sec:conc}
We have presented the first public annotated infant respiration dataset, AIR-125, together with a novel deep learning model, AIRFlowNet, tuned for infant subjects and achieving state-of-the-art performance on AIR-125. Our model uses optical flow, spatio-temporal learning, and a new spectral bandpass loss function to optimize performance across varied lighting and camera settings, toward the eventual goal of automated, continuous, and purely video-based infant respiratory monitoring, both in NICU and at-home settings. Fruitful work in the future could include expanding the dataset scope and model performance, or achieving similar performance without using dense optical flow to enable immediate real-time monitoring in critical care situations.


\bibliographystyle{splncs04}
\bibliography{biblio}

\end{document}